\documentclass[journal]{vgtc}                % final (journal style)
\ifpdf%                                % if we use pdflatex
  \pdfoutput=1\relax                   % create PDFs from pdfLaTeX
  \usepackage{graphicx}                % allow us to embed graphics files
  \DeclareGraphicsExtensions{.pdf,.png,.jpg,.jpeg} % for pdflatex we expect .pdf, .png, or .jpg files
\else%                                 % else we use pure latex
  \usepackage{graphicx}                % allow us to embed graphics files
  \DeclareGraphicsExtensions{.eps}     % for pure latex we expect eps files
\fi%

\graphicspath{{figures/}{pictures/}{images/}{./}} % where to search for the images

\usepackage{microtype}                 % use micro-typography (slightly more compact, better to read)
\PassOptionsToPackage{warn}{textcomp}  % to address font issues with \textrightarrow
\usepackage{textcomp}                  % use better special symbols
\usepackage{mathptmx}                  % use matching math font
\usepackage{times}                     % we use Times as the main font
\usepackage{cite}                      % needed to automatically sort the references
\usepackage{tabu}                      % only used for the table example
\usepackage{booktabs}                  % only used for the table example
\usepackage{paralist}
\usepackage{color}
\usepackage{tabularx}
\usepackage{hyperref}
\usepackage{amsmath}
\usepackage{amsfonts}
\usepackage{tabularx}
\usepackage{multirow}

\usepackage{tabularx}
\usepackage{cases}
\usepackage{setspace}

\DeclareMathAlphabet{\mathcal}{OMS}{cmsy}{m}{n}

\newcommand{\system}{DashBot}
\newcommand{\vslen}{-8px}
\newcommand{\vplen}{-8px}
\newcommand{\splen}{0.97}

\usepackage[dvipsnames]{xcolor}

\onlineid{1033}

\vgtccategory{Research}
\vgtcpapertype{Algorithm (Area 5: Data Transformations)}

\title{\system{}: Insight-Driven Dashboard Generation Based on\\Deep Reinforcement Learning}

\author{Dazhen Deng, Aoyu Wu, Huamin Qu, and Yingcai Wu}
\authorfooter{
\item D. Deng and Y. Wu are with the State Key Lab of CAD\&CG, Zhejiang University. Y. Wu is also with the Alibaba-Zhejiang University Joint Research Institute of Frontier Technologies. E-mail: \{dengdazhen, ycwu\}@zju.edu.cn.
\item A. Wu and H. Qu are with Department of Computer Science and Engineering, the Hong Kong University of Science and Technology. E-mail: \{awuac, huamin\}@cse.ust.hk.
\item Yingcai Wu is the corresponding author.
}

\abstract{Analytical dashboards are popular in business intelligence to facilitate insight discovery with multiple charts. However, creating an effective dashboard is highly demanding, which requires users to have adequate data analysis background and be familiar with professional tools, such as Power BI. To create a dashboard, users have to configure charts by selecting data columns and exploring different chart combinations to optimize the communication of insights, which is trial-and-error. Recent research has started to use deep learning methods for dashboard generation to lower the burden of visualization creation. However, such efforts are greatly hindered by the lack of large-scale and high-quality datasets of dashboards. In this work, we propose using deep reinforcement learning to generate analytical dashboards that can use well-established visualization knowledge and the estimation capacity of reinforcement learning. Specifically, we use visualization knowledge to construct a training environment and rewards for agents to explore and imitate human exploration behavior with a well-designed agent network. The usefulness of the deep reinforcement learning model is demonstrated through ablation studies and user studies. In conclusion, our work opens up new opportunities to develop effective ML-based visualization recommenders without beforehand training datasets.} % end of abstract

\keywords{Reinforcement Learning, Visualization Recommendation, Multiple-View Visualization}

\CCScatlist{ % not used in journal version
 \CCScat{K.6.1}{Management of Computing and Information Systems}%
{Project and People Management}{Life Cycle};
 \CCScat{K.7.m}{The Computing Profession}{Miscellaneous}{Ethics}
}

\teaser{
  \centering
  \includegraphics[width=0.99\linewidth]{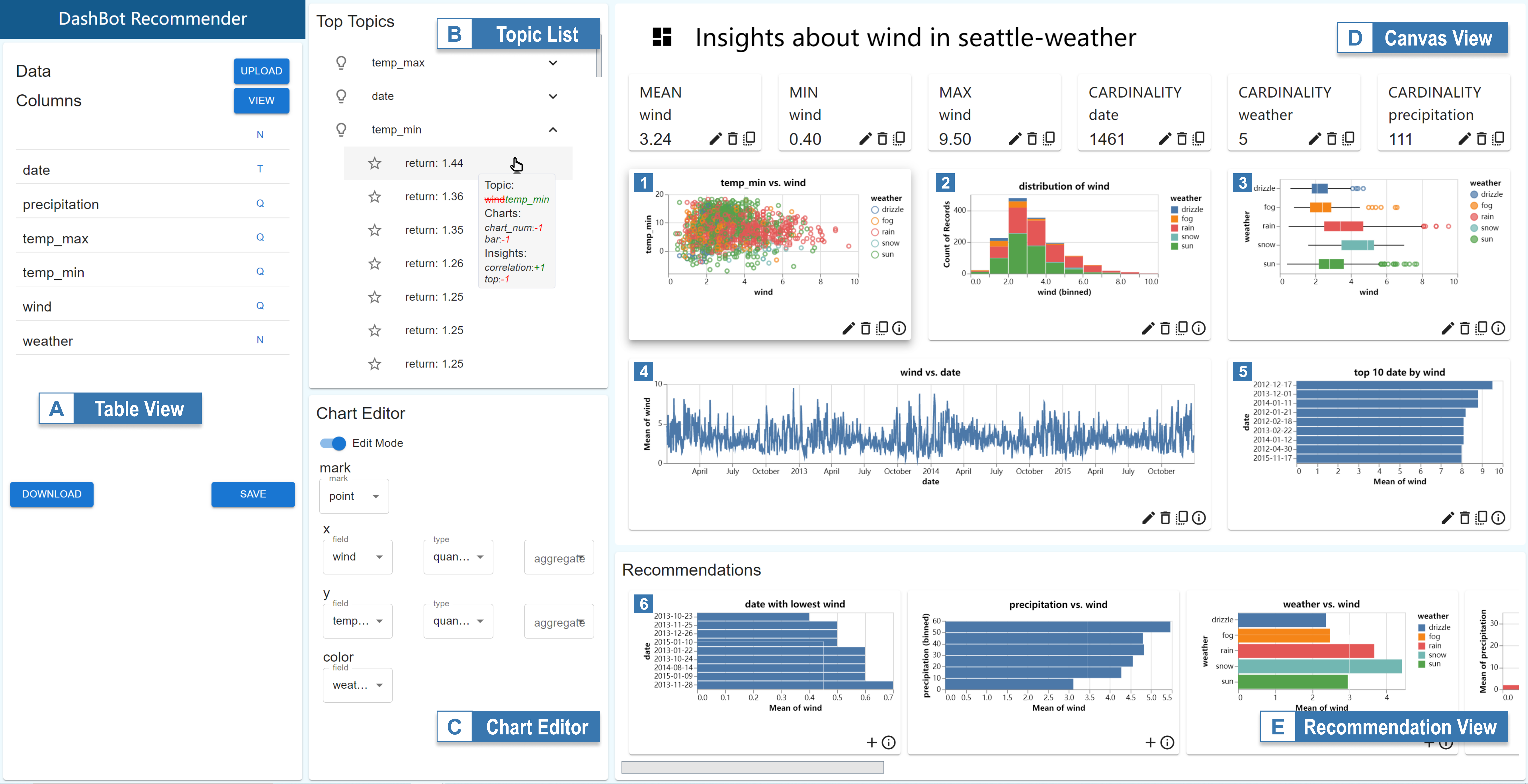}
	\vspace{\vslen}
  \caption{Screenshot of the \system{} Interface showing the generation of a dashboard named ``Insights about wind in seattle-weather''. The interface consists of a table view (A), a topic list (B), a chart editor (C), a canvas view (D), and a recommendation view (E).}
  \label{fig:teaser}
}

\vgtcinsertpkg

\begin{document}

%% The ``\maketitle'' command must be the first command after the
%% ``\begin{document}'' command. It prepares and prints the title block.

%% the only exception to this rule is the \firstsection command

\maketitle

\begin{spacing}{\splen}
\section{Introduction}
% Analytical dashboards are important but hard to generate
Analytical dashboards have been broadly used in business intelligence to help data analysts explore and discover data insights with multiple-view visualizations (MVs)~\cite{sarikaya2018we}. 
Even with the help of professional authoring tools, such as Tableau and Power BI, creating an effective dashboard is still a highly demanding task, requiring expertise in data analysis and visualization. 
Specifically, the analyst needs to explore the dataset, select appropriate data columns and visual encodings to configure charts, and investigate whether the charts are insightful. 
In addition, the analyst has to consider the relationship between charts to exhibit different perspectives of the dataset~\cite{wang2000guidelines}.
Such a process of exploratory data analysis for dashboard generation is trial-and-error~\cite{battle2019characterizing}.
% \dazhen{highlight the EVA}

To reduce the burden, many studies have investigated rule-based and machine learning-based (ML-based) methods for visualization recommendation.
Rule-based methods, such as APT~\cite{mackinlay1987automatic}, CompassQL~\cite{wongsuphasawat2016towards}, and Voyager~\cite{wongsuphasawat2015voyager, wongsuphasawat2017voyager}, translate well-established visualization design rules (e.g., expressiveness and effectiveness criteria~\cite{mackinlay1987automatic}) to be programmable constraints for the recommendation.
Differently, ML-based methods employ state-of-the-art models, such as decision trees ~\cite{luo2018deepeye} and deep neural networks~\cite{li2021kg4vis, dibia2019data2vis, hu2019vizml}, to learn common patterns of visual encodings from large-scale chart datasets~\cite{hu2019viznet, deng2020visimages}.
Though useful in generating effective charts, these methods focus on single visualizations instead of MVs, where the relations between charts are important.
% in which additional problems such as view selection~\cite{wang2000guidelines} and view consistency~\cite{qu2017keeping} should be addressed.

Towards the generation of MVs, a series of studies investigate the use of hand-crafted rules.
For example, to tell data stories, Datashot~\cite{wang2019datashot} and Calliope~\cite{shi2020calliope} adopt statistics metrics (e.g., Pearson correlation coefficient) to extract facts from datasets and then generate charts according to the facts.
These successful cases prove the usefulness of hand-crafted rules.
Recently, deep learning-based methods have also been used to improve the efficiency of generating MVs.
For example, MultiVision~\cite{wu2021multivision} trains deep neural networks to score the goodness of single charts.
Then the single chart scoring is combined with customized metrics to generate multiple charts.
In this method, the generation of MVs is conducted indirectly due to the lack of high-quality MVs datasets, 
which could harm the training process and results.
% which might hinder the development of deep learning-based approaches.

In this paper, we propose to use deep reinforcement learning to generate analytical dashboards, taking advantage of established visualization knowledge and efficient machine intelligence. 
We argue that applying reinforcement learning to such a scenario has several advantages.
1) Intuitive modeling. Reinforcement learning agents learn from environments through exploration, which is similar to the mechanism of exploratory visual analysis. Therefore, modeling the exploratory visual analysis process with reinforcement learning is natural.
2) Self-play training. With carefully designed environments, action spaces, and reward functions, agents can be constantly trained with different datasets and obtain a shared experience of dashboard generation, with no need for beforehand training data with labels.
3) Online recommendation. Due to a similar mechanism, a reinforcement learning-based recommendation system supports a swift switching between human steering and automation. When users update the dashboards by preference, the agents can generate subsequent recommendations promptly.

However, designing a reinforcement learning model for analytical dashboards is challenging.
First, it lacks a well-established environment for ``visualization agents'' to explore and train.
Different from training agents to play games (e.g., AlphaGo~\cite{silver2016mastering}), in which the winning and losing can explicitly measure the goodness of agent actions, there is no deterministic rule to evaluate the generated dashboards.
Second, it is difficult to design an agent to imitate complex human behavior in generating dashboards, including configuring charts by multiple parameters (e.g., mark types, encoding types, data fields, and data transformations) and exploring the large space of chart combinations. 

To address the above challenges, we developed \system{}, a deep reinforcement learning-based recommendation system for analytical dashboards.
For the first challenge, we investigated state-of-the-art design guidelines for MVs~\cite{wang2000guidelines, qu2017keeping} and conducted a preliminary study with a collection of dashboard designs of Tableau and Power BI.
From the study, we investigated how to use design guidelines for assessing dashboards.
Built upon the guidelines and knowledge gained, we designed a dashboard playground, which scores the generated dashboards and provides an interface for the agents to explore the dashboard design space.
For the second challenge, we formulated the exploratory dashboard generation to be a sequence prediction problem.
Specifically, at each state, agents can decide the actions to take (e.g., adding or removing a chart) and configure chart parameters (e.g., chart types and encoding types).
We designed a novel deep neural network to achieve the prediction.
To improve training efficiency, we proposed a constrained sampling strategy to ensure the validity of generated charts while preserving the exploration uncertainty of the agents.
To validate the usefulness of the proposed method, we conducted an ablation study for the model design and comparative experiments with a state-of-the-art dashboard generation system.
We also analyzed the user feedback and reflected on designing automatic agents to generate analytical dashboards.
In summary, we have four major contributions.
\begin{compactitem}
    \item A preliminary study that reviews practical dashboard designs and summarizes the design considerations for the recommendation.
    \item A reinforcement learning formulation for dashboard generation that features the definition of reward functions for evaluating the expressiveness and insightfulness of the dashboards.
    \item A novel deep neural network for agents to explore the actions and parameters for generating dashboards.
    \item A series of quantitative and qualitative studies that validate the usefulness of the proposed approach and lessons learned in designing automatic agents for visualizations.
\end{compactitem}

\section{Related Work}
In this section, we introduce related studies from the perspectives of visualization recommendation, multiple-view visualization generation, and reinforcement learning for visualization.

\subsection{Visualization Recommendation}
Existing visualization recommendation approaches can be categorized into rule-based methods and ML-based methods~\cite{saket2018beyond, wu2021survey}.
Rule-based methods utilize the principles in visualization theories to construct visual mapping.
For example, APT~\cite{mackinlay1987automatic} incorporates expressiveness and effectiveness criteria~\cite{bertin1983semiology} into graphical languages to formulate visualizations.
Show Me~\cite{mackinlay2007show} and CompassQL~\cite{wongsuphasawat2016towards} employ query techniques to enumerate visual encodings.
Furthermore, Voyager~\cite{wongsuphasawat2015voyager, wongsuphasawat2017voyager} adopts statistics and perceptual measures to rank the generated visualizations and supports interactive exploration.

ML-based methods incorporate machine learning models to predict the visual mapping.
A number of methods formulate the visual mapping as a non-linear regression from hand-crafted data features to charts, such as VizML~\cite{hu2019vizml}, NL4DV~\cite{narechania2020nl4dv}, and wide-and-deep recommendation network~\cite{qian2020ml}.
Other methods formulate the recommendation as different problems, such as sequence-to-sequence translation~\cite{dibia2019data2vis,zhou2020table2charts}, learning-to-rank~\cite{luo2018deepeye, wu2021learning,moritz2018formalizing, sun2022learning}, 
and knowledge graph~\cite{li2021kg4vis, zhu2021visualizing}.
However, these recommendation methods mainly focus on generating a single chart, which might be insufficient for solving the visual analysis problems with high-dimensional data.

\subsection{Multiple-View Visualization Generation}
Multiple-view visualizations (MVs) are useful in visual analysis for their capability in representing different perspectives of data simultaneously.
Numerous MVs, which refer to visual analytics (VA) systems, have been created to discover patterns and insights~\cite{andrienko2021theoretical, wu2022defence}.
Existing studies of visualization recommendation for VA systems focus on layout problems regarding organizing multiple views~\cite{shao2021modeling}.
For example, Al-maneea and Roberts~\cite{almaneea2019quantifying} proposed a series of criteria to decompose the VA systems in the publications and quantify their layouts.
Chen et al.~\cite{chen2020composition} investigated view composition and configuration of the systems.
% Furthermore, they developed an interface for view recommendation based on the view position and types.
% However, there is still a long way to go for automatic VA system ``generation'' because of the complex configurations of data and charts.

Existing studies of MV generation target to creating MVs from tabular data for insight discovery~\cite{karer2021insight} or storytelling~\cite{segel2010narrative, edmond2021three}.
For example, Voder~\cite{srinivasan2018augmenting} and QRec-NLI~\cite{wang2022interactive} adopts natural language processing models to extract data facts or recommending next-step queries for dashboard exploration.
Zhao et al.~\cite{zhao2021chartstory} proposed ChartStory, a system that composes charts into comic-style visualizations.
DataShot~\cite{wang2019datashot} generates data fact sheets with a template-based method for visual storytelling.
Similarly, Calliope~\cite{shi2020calliope} obtains data insights with customized metrics and identifies the best ones using a Monte Carlo tree search.
These rule-based methods can well integrate visualization domain knowledge into the design of metrics.
Recent research also explores the generation of dashboards with deep learning.
MultiVision~\cite{wu2021multivision} employs bidirectional long short-term memory models to score and rank single charts for MV generation.
In this work, we integrate deep learning methods and visualization knowledge to generate analytical dashboards.
Specifically, we utilize the capability of deep neural networks in simulating complex environments and take advantage of well-established visualization design rules to score the generated visualizations.

\subsection{Reinforcement Learning and Visualization}
Reinforcement learning aims to train agents to take actions in specific environments so that the agents can gain the highest accumulated rewards.
Given a current observation, Q-learning~\cite{watkins1992q} is designed to predict the rewards that can be gained and choose the actions with the highest rewards.
However, Q-learning can only handle a small number of observations and actions.
To cope with the problems with complex situations, more variants based on deep learning have been proposed.
For example, to handle the large observation space of game screens during playing Atari video games, Mnih et al.~\cite{mnih2013playing} proposed deep Q-learning.
Though useful, deep Q-learning fails when there is a high-dimensional action space.
The problem of instability during training also arises.
A series of policy gradient methods~\cite{lillicrap2015continuous, schulman2015trust, schulman2017proximal, mnih2016asynchronous} have been proposed to address these problems.
In this work, we choose to use asynchronous advantage actor-critic algorithm (A3C)~\cite{mnih2016asynchronous}, a state-of-the-art reinforcement learning framework, to handle the problem of dashboard generation, in which both observation space (i.e., dashboards with different chart combinations and variant chart numbers) and action space (i.e., generating chart configurations) are high-dimensional.

A few studies have used reinforcement learning models for visualization generation.
For example, MobileVisFixer~\cite{wu2020mobilevisfixer} adopts an explainable Markov decision model to optimize the layouts of visualizations on mobile devices.
Bako et al.~\cite{bako2021user} also adopted a Markov decision process model to recommend potential D3 syntax for authoring visualizations.
Tang et al.~\cite{tang2020plotthread} adopted reinforcement learning to create storyline layouts.
Shi et al.~\cite{shi2019task} and Wei et al.~\cite{wei2022evolutional} used reinforcement learning to predict next-step operations of chart editing.
In this work, we target the dashboard, a multiple-view visualization with larger design space, and formulate the problem of dashboard generation to be a reinforcement learning problem.

\section{Design of \system{}}
To design an effective recommendation system, we conducted a preliminary study to understand the current practices of analytical dashboards and derive the design considerations of recommendation systems.

\subsection{Preliminary Study}
Existing studies for visualization designs mainly focus on visualization genres such as infographics~\cite{borkin2013makes}, storylines~\cite{tang2018istoryline}, and data stories~\cite{wang2019datashot, shu2020makes}.
There are few studies that investigate the design patterns for analytical dashboards~\cite{sarikaya2018we, bach2022dashboard}.
Therefore, we start with a preliminary study to gain an overview of the design practices. 

\begin{figure}[tb]
	\includegraphics[width=\linewidth]{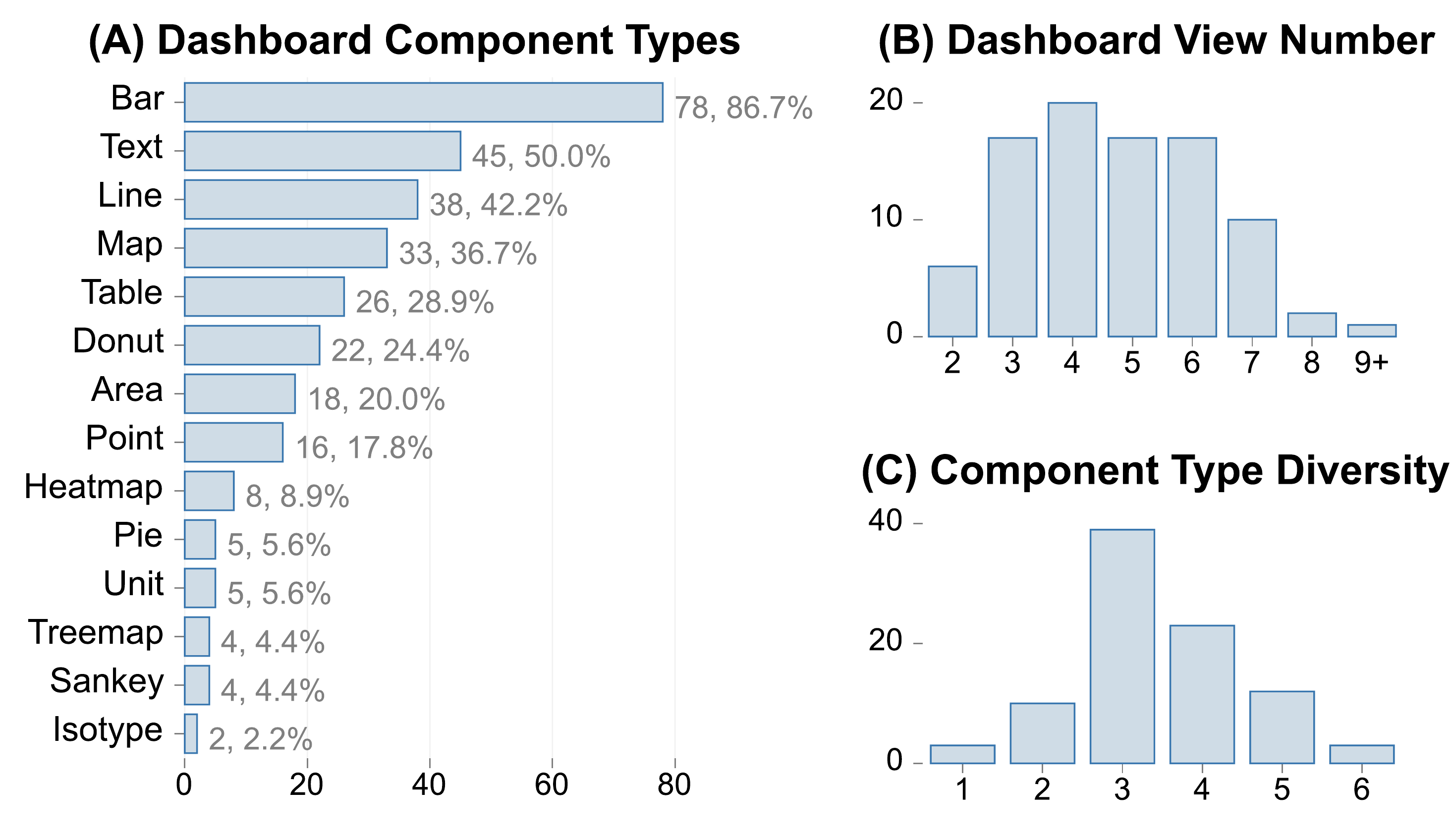}
	\vspace{\vslen}
	\caption{Statistics in the preliminary study: distributions of dashboard component types, numbers of views in a dashboard, and numbers of different component types in a dashboard.}
	\vspace{\vplen}
	\label{fig:preliminary}
\end{figure}
% \subsection{Data Collection}

We first collected dashboards with the most number of ``likes'' from the official galleries of Tableau~\cite{tableauGallery} and Microsoft Power BI~\cite{powerbiGallery}, which contain hundreds of high-quality examples.
Not all examples in the galleries are dashboards and some of them are posters (telling stories mainly with text and charts only for assistance) or infographics (with only one well-designed view).
Therefore, we carefully filtered the dashboards from the galleries.
% In addition, if a dashboard contains multiple pages, we divided it by pages and analyzed each page.
Finally, we obtained 40 Tableau dashboards and 50 Power BI dashboards.

Second, we analyzed the dashboards from visual and data presentation perspectives.
Based on the design guidelines of multiple-view visualizations~\cite{wang2000guidelines}, a good design should follow the rules of diversity, complementarity, decomposition, and parsimony. 
We opted to understand the collected designs concerning these rules.
% However, the definitions of the rules are ambiguous, lacking concrete metric to assess the design examples. 
% Therefore, we derive practical metrics following the ideas of the rules and present our analysis results as follows. 

\textbf{Diversity.} We analyzed how diverse chart types are used to represent the data columns. A dashboard usually contains multiple views, and each comprises one or multiple charts or components (e.g., text and table). Therefore, for each dashboard, we annotated the types of charts and components, as shown in \autoref{fig:preliminary}(A). From the results, we discovered that the bar chart is the most commonly used chart type, followed by line charts, maps, and donut charts. 
Text components are the second most common, used to summarize the insights in the charts or show key indicators independently.
The table components are usually used to show raw data directly. 

\textbf{Parsimony.} 
The rule of parsimony refers to minimizing the number of views while preserving effectiveness and expressiveness. Therefore, we counted the view numbers of the collected dashboards (\autoref{fig:preliminary}(C)). We discovered that most dashboards are composed of 3-6 views. Only a small number of dashboards contain more than 8 views.

\textbf{Complementarity \& Decomposition.} Complementarity refers to how charts complement each other to exhibit different perspectives of the datasets. 
Based on the definition in the prior study~\cite{wu2021multivision}, we regard two views to be complementary to each other when they visualize different data columns. 
For example, a view encodes columns A and B and another view encodes columns C and D. 
On the contrary, decomposition refers to analyzing complex data with multiple charts, such as chunking the data or applying different aggregation methods to a data column. These charts will share the same data column.
When investigating the examples in-depth, we discovered that few dashboards include views complementary to each other, because the dashboards usually concern specific ``key columns''. In fact, in most dashboards (96.7\%), all their views are related to one or two data columns. 

% \textbf{Decomposition.} In comparison with complementarity, decomposition refers to analyze complex data with multiple charts, such as chunking the data or applying different aggregation methods on a data column.
% In this work, we regard two views to be decomposed to each other when they share a same data column. For example, a view encode columns A and B and another view encodes columns A and C. 
% Based on the analysis of complementarity, in most dashboards, views are combined together to decompose the problem of analyzing the key columns.

From the analysis above, we understood that to design an effective dashboard, it is encouraged to identify a topic (i.e., a key column) and configure composed charts to discover insights surrounding the topic~\cite{wang2019datashot}.
Besides, it is necessary to introduce adequate chart diversity to enhance expressiveness but avoid a large chart number.

% It is noted that there are additional rules for user interactions and visual styles that have been studied by existing research~\cite{qu2017keeping}.
% We currently focus on data insight presentation, which is the core task for analytical dashboards.

\subsection{Design Considerations}
% We develop a mixed-initiative recommendation system for analytical dashboards.
Based on previous empirical studies~\cite{wang2000guidelines, qu2017keeping}, recommendation systems~\cite{shi2020calliope, wu2021multivision}, and our preliminary study on current practices, we derive design considerations of a recommendation system for dashboards.
\begin{figure*}[tb]
	\includegraphics[width=\linewidth]{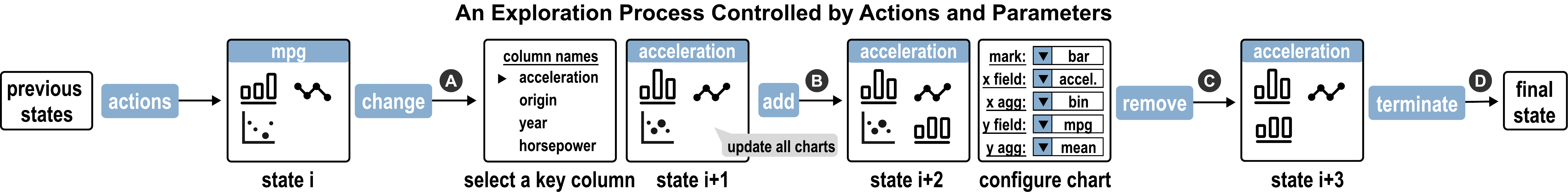}
	\vspace{\vslen}
	\caption{An exploration process is controlled by actions and parameters. The agent first changes the key column at state i (A) and adds a new chart into the dashboard by configuring chart parameters (B). Then the agent removes a chart (C) and terminates the whole exploration process (D).}
	\vspace{\vplen}
	\label{fig:actions}
\end{figure*}
\begin{compactenum}[\bfseries DC1]
\item \textbf{Generate valid dashboards automatically.} The dashboards should be automatically generated with specific characteristics. Based on the preliminary study, an analytical dashboard commonly features multiple charts with diverse types and components of text and tables. The charts should also follow effectiveness rules~\cite{mackinlay1987automatic}. For example, bar charts are suitable to visualize the data of a nominal column and a quantitative column. 
\item \textbf{Facilitate self-steering data insight discovery.} In addition to merely visualizing the data with appropriate visual encodings, an effective dashboard is supposed to convey data insights. When generating the dashboard, the system should recognize data insights, such as high correlations and temporal distributions, and prioritize selecting charts with insights into the dashboard.
\item \textbf{Enable direct manipulation on the recommendations.} It is unlikely to generate dashboards that fulfill the requirements of all users. Therefore, users should be allowed to modify the dashboards directly. The system should provide an interface for users to customize the dashboards according to their preferences and add new charts in an exploratory manner.
\item \textbf{Support online recommendation during exploration.} Editions on the dashboard inherently exhibit user preferences, which provide additional conditions for the recommendation.
Therefore, the system should be able to start from current dashboard configurations and explore the best dashboards accordingly.
Furthermore, the system should promptly generate new recommendations after the modifications to ensure interaction efficiency.
% \item \textbf{Facilitate convenient result exporting and sharing.}
\end{compactenum}
Guided by these design considerations, we develop a recommendation system empowered by a deep reinforcement learning-based computation module.
The computation module is built with a deep neural network that can generate valid charts (\textbf{DC1}) and discover data insights (\textbf{DC2}) automatically. To ensure the validity of the generated charts, we propose a novel constrained sampling method to apply rules to the sampling state (\textbf{DC1}). Besides, we carefully design insight rewards to encourage the discovery of insights during the exploration (\textbf{DC2}).
On top of the computation module, we develop an interactive interface that allows users to edit the recommended dashboards (\textbf{DC3}).
The user editions are then fed back to the computation module, and the module generates new recommendations based on the editions (\textbf{DC4}).

\section{Problem Formulation}
This section introduces how to formulate exploratory dashboard generation to be a reinforcement learning problem.
We first introduce the Markov decision process (MDP), the foundation of reinforcement learning. Then we interpret human behaviors of designing analytical dashboards into MDP actions.
\subsection{Background: Markov Decision Process}
Markov decision process (MDP)~\cite{bellman1957markovian} is a stochastic control process.
An MDP can be formulated as a sequence of states and actions:
\begin{equation}
P=\{(s_i, a_i, p_i, r_i)|i\in[1,T]\},
\end{equation}
where $T$ is the length of the process, $s_i\in\mathcal{S}$ is the state at step $i$, $a_i\in\mathcal{A}$ is a selected action from action space $\mathcal{A}$, $p_i\in\mathbb{R}^{|\mathcal{A}|}$ is the probabilities of different actions, and $r_i\in\mathbb{R}$ is the immediate reward after executing action $a_i$ at state $s_i$.

% In an MDP, the agent receives an immediate reward measuring its action at each state.
The key of reinforcement learning (RL) is to train agents obtain the most cumulative rewards after the exploration, i.e., return $R=\sum_{i=1}^{T} \gamma^{i-1} r_i$, where $r_i$ is the immediate reward obtained at each state and $\gamma$ ($\gamma=1$) is a discount rate.
Therefore, the MDP is suitable to model exploratory processes, in which the goodness of the actions for the final result can be quantitatively evaluated at each state.

\subsection{Formulating Dashboard Generation as MDP}
\label{sec:mdp}
We define the states and actions of designing an analytical dashboard. Based on the preliminary study, we model the generation of dashboards as a process to identify a key column (i.e., topic) from the dataset and configure a series of charts with additional explanation columns to reveal insights about the topic.

\textbf{State Space.} During dashboard generation, the agent decides the next actions according to the current configuration of the dashboard, which is a collection of charts.
Therefore, the state space $\mathcal{S}$ enumerates all valid chart combinations with a given dataset.
\begin{equation}
    \mathcal{S}=\big\{\{chart_{j}|j\in[0,n]\}\big|n\in[0,N]\big\},
\end{equation}
where $chart_j$ denotes a chart and $N$ is the maximum chart number.

% \begin{table}[tb]
%     \small
%     \caption{Actions and Parameters.}
%     \begin{tabularx}{\linewidth} {XXX}
%         \toprule
%         \bf Actions & \bf Parameters & \bf Descriptions\\\midrule
%         change & column & Select a column as key column from the dataset.\\\midrule
%         add & chart types and visual encodings (x field, x aggregation, y field, etc.) & Configure a new chart by chart types and visual encodings and add it to the dashboard.\\\midrule
%         remove & chart to remove & Remove a chart from the dashboard\\\midrule
%         terminate & - & End the exploration session and submit the generated dashboard.\\\bottomrule 
% \end{tabularx}
% \label{tab:actions}
% \end{table}

\textbf{Action Space.} From a specific state, the agent continues to explore the dashboard design space by taking action. To fully explore the design space, agents are allowed to \textit{change} the key column, \textit{add} a new chart, \textit{remove} an existing chart, and decide whether to \textit{terminate} the exploration session.
The action space is defined as follows.
\begin{equation}
    \mathcal{A}=\{change, add, remove, terminate\}
\end{equation}

Different actions require different parameters for execution.
{When the agent decides to \textit{change} the key column, it should further select a new one from the dataset. All charts will replace the key column with the new one.}
To \textit{add} a new chart, the agent has to specify mark types and visual encodings for the chart.
The action \textit{remove} refers to removing an existing chart from the dashboard.
The action \textit{terminate} will end up the exploration session and finalize a dashboard.
An example of executing actions with parameters is demonstrated in \autoref{fig:actions}.
% As suggested by Bar et al.~\cite{bar2020automatically}, sub-modules followed by action predictions can be introduced to handle the parameter selections for different actions.
% Existing deep learning methods for visualization configuration~\cite{hu2019vizml, dibia2019data2vis,luo2021natural} usually adopt multi-classification neural networks.
% Inspired by these methods, we predict the values of mark types and visual encodings after deciding to add a new chart.
% Existing studies have investigated the chart configuration space.
% For example, Kim et al.~\cite{kim2017graphscape} formulate the generation of charts as a directed graph model. Given a chart, users can generate a new one by editing the mark type, transforming data columns, or modifying the visual encoding.
% This formulation results in the generation of chart sequences in which neighboring charts changes incrementally~\cite{shi2019task}.
% Considering the scenario of dashboard generation, the creation of one valid chart might be composed of a long sequence of operations.
% Another is totally excluding the chart configuration during the MVs generation.
% For example, Shi et al.~\cite{shi2020calliope} first extracted data insights and configured the charts for the insights with default settings. Visual encodings are not considered in their proposed method.
% In this work, to facilitate efficient dashboard generation, we allow agents to explore the visual encodings and column selections to an adequate degree.

\subsection{Reward Functions}
After executing the action and obtaining a dashboard with a new state, the environment is supposed to return a reward to tell the agent how well is the generated dashboard from the previous action. 
Reward functions are critical for training effective agents to achieve the goals, which should be carefully designed based on existing knowledge about dashboard design.
The immediate reward at step $i$ is defined as follows,
\begin{equation}
    r_{i}=f(s_{i},a_{i}),
\end{equation}
which is computed from the action $a_i$ and state $s_i$.
We have designed two categories of reward functions, namely, presentation rewards and insight rewards.
\subsubsection{Presentation Rewards}
The presentation rewards are derived from the rules of diversity and parsimony based on the preliminary study.

\textbf{Diversity Reward.} Dashboards commonly adopt different chart types to improve expressiveness. However, a higher diversity is not always better because users have to interpret the visual encodings of different charts, increasing users' cognition load~\cite{qu2017keeping}.
From the preliminary study, most of the dashboards contain 2 or 3 different chart types. 
Therefore, we use a concave increasing function to award the increase of diversity with the idea of ``diminishing returns'':
\begin{equation}
    dr(s_i, a_i)=1-exp(-\frac{\alpha\cdot c_{used}}{c_{total}}).
\end{equation}
The variable $c_{used}$ denotes the number of chart types used in the dashboard and $c_{total}$ is the total number of chart types allowed, and $\alpha$ is a weight to adjust the diminishing degree. Similarly, we use the function to score the diversity of the visualized columns with regard to all columns of the dataset.

\textbf{Parsimony Reward.} Similar to chart diversity, the chart numbers should not be too large. From the preliminary study, most of the dashboards contain 3-5 different charts. However, different from chart types that have limited choices, the number of charts can be ambiguously large. Therefore, the parsimony reward is a piece-wise function that firstly increases and then decreases with regard to the chart numbers:
% \begin{numcases}{parsimony(s_i)=}
%     \sin{\frac{\pi}{2}\cdot\frac{n}{n_{best}}},  & $n \in[0,n_{best}]$
%    \\
%    \sin{\frac{\pi}{2}\cdot(1+\frac{n-n_{best}}{n_{max}-n_{best}})},& $n \in(n_{best}, n_{max}]$ 
% \end{numcases}
\begin{equation}
\begin{aligned}
    pr(s_i, a_i)= &
     \begin{cases}
        \sin{\frac{\pi}{2}\cdot\frac{n}{n_{best}}},        & n \in[0,n_{best}] \\
        \sin{\frac{\pi}{2}\cdot(1+\frac{n-n_{best}}{n_{max}-n_{best}})},        & n \in(n_{best}, n_{max}].
     \end{cases}
     \\
   \end{aligned}
\end{equation}
The variable $n$ is the chart number of the dashboard, $n_{best}$ is the best number of charts, and $n_{max}$ is the maximum number of charts.

\subsubsection{Insight Rewards}
The use of a dashboard is to identify and visualize insights.
Therefore, we design insight rewards to award the discovery of insightful charts.
We enumerate the metrics proposed by existing methods~\cite{shi2020calliope, wang2019datashot, bar2020automatically, tang2017extracting} and categorize the insight metrics based on the number of columns involved, including single-column insights and double-column insights.
Additionally, we propose to measure multiple-column insights on the basis of single and double-column insights.

Concerning a single column, it is common to exhibit the statistics (e.g., mean and cardinality) or value distribution.
These insights can give an overview of the selected column, but users might be more interested in the relation between the two columns.
Previous studies~\cite{shi2020calliope, wang2019datashot} have adopted statistical metrics to measure double-column insights, such as the Pearson correlation coefficient.
In this work, we model the double-column insights of a dashboard in a similar way.
% For example, if the agent visualizes two columns with higher correlation, it will receive a higher reward.

Furthermore, we propose to model multiple-column insights for dashboard insight discovery. 
The multiple-column insights refer to the insights derived from multiple charts, where more than two explanation columns are involved.
For instance, if column A has a high correlation with column B and column A has a high correlation with column C, it is possible that column B correlates to column C.
Existing methods mainly focus on the insights in single charts~\cite{wang2019datashot} or relations between two neighboring charts~\cite{shi2020calliope}.
Differently, we model the dashboards as a collection of charts and consider the relations between any two charts.

% Definitions of insights are demonstrated in \autoref{tab:insight_reward}.
For each insight, we demonstrate the conditions for chart types and column types and the statistical conditions to be fulfilled (\autoref{tab:insight_reward}).
When recognizing a single/double/multiple-column insight, the reward value would be $1$, $2$, and $3$, respectively.

\subsubsection{Combined Rewards}
We combine the rewards together by weighted sum:
\begin{equation}
    \textstyle cr_{i}=w_1\cdot\sum_{c}^{\{col, vis\}} dr^c_i+w_2\cdot pr_i+ w_3\cdot\sum_{c}^{\{insights\}} ir^c_i,
\end{equation}
where $\{w_k\}$ are constant values that balance the magnitude of the rewards.
We empirically set $w_1=w_2=0.33$ to normalize the presentation rewards and set $w_3=0.1$ to encourage the gaining of the insights.
We compute the immediate reward gained at state $i$ by
\begin{equation}
    f(s_{i},a_{i})=cr_i-cr_{i-1},
\end{equation}
where $cr_i$ and $cr_{i-1}$ are dashboard rewards at states $i$ and $i-1$.

\begin{table}[tb]
    \centering
    \caption{Definitions of the Insights.}
    % \normalsize
    \small
    \begin{tabularx}{\linewidth}{lX}
    \toprule
        \textbf{Insight} &\textbf{Definition}\\\midrule
        % cardinality & $\exists chart(A)\in\{text\}, A\in\mathcal{N}:|A|>thre.$\\\midrule
        distribution & $A\in\mathcal{Q}$: visualize A with a histogram by applying bin count.\\\midrule
        trend & $A\in\mathcal{Q}, B\in\mathcal{T}$: visualize A across B with a line chart.\\\midrule
        correlation &$A\in\mathcal{Q}, B\in\mathcal{Q}$: visualize A across B with a line chart or scatterplot, and the correlation between A and B is higher than the threshold.\\\midrule
        top/bottom k & $A\in\mathcal{N}, B\in\mathcal{Q}$: visualize top or bottom k entities of A with B.\\\midrule
        co-correlation & $A\in\mathcal{Q}, B\in\mathcal{Q}, C\in\mathcal{Q}$: there are correlation insights about (A, B) and (A, C).\\\midrule
        comparison & $A\in\mathcal{N}, B\in\mathcal{Q}$: there are top and bottom k insights about A and B.\\\bottomrule
        \multicolumn{2}{p{\linewidth}}{*note: $\mathcal{Q}$, $\mathcal{T}$, and $\mathcal{N}$ stand for quantitative, temporal, and nominal columns.}
    \end{tabularx}
    \vspace{\vplen}
    \label{tab:insight_reward}
\end{table}

\section{Design of \system{}}
Based on the formulation of dashboard generation, we further design proper machine learning models to facilitate the exploration.

\subsection{Model Framework}

In this work, we use the asynchronous advantage actor-critic algorithm (A3C)~\cite{mnih2016asynchronous}, a novel reinforcement learning framework, to train agents to explore the dashboard design space.
A3C is a flexible framework designed for the problem with a large action and observation space, which is well-suited for the problem of visualization generation~\cite{tang2020plotthread}.

% In addition, it is difficult to tell which actions contribute more for the return.
% Advantage actor-critic algorithm (A3C) is a recently proposed method to handle the problem with large action space.
A3C framework consists of two modules, an action network predicting the action probabilities and a critic network evaluating the maximum expected return from the state.
The main idea of A3C is to learn an accurate critic network to evaluate the expected return from the state.
Then the critic network is used to train the action network to take the best actions from specific states.
During training, the return $R$, probabilities of the actions $p_{i}$, value estimation for the expected return $v(s_i)$, and rewards $r_i$ are fed into a loss function:
\begin{equation}
	L(s_i,p_i) = (R-v(s_i))^2-\log(p_i)A(s_i)-H(p_i),
\end{equation}
where $A(s_i)$ is advantage function and $H(p_i)$ is entropy of the action probabilities~\cite{mnih2016asynchronous}.
The framework of the A3C is shown below.

\begin{figure}[!h]
	\includegraphics[width=\linewidth]{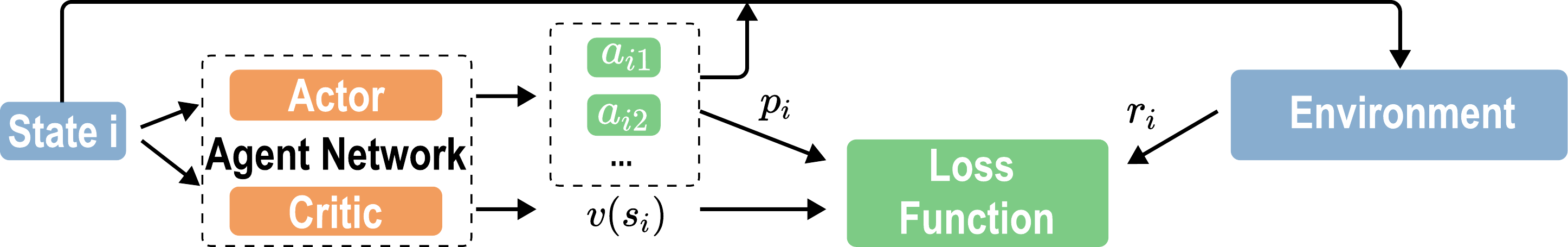}
	\vspace{\vplen}
	% \caption{Framework of advantage actor-critic algorithm.}
	% \label{fig:a3c}
\end{figure}

\begin{figure*}[tb]
	\includegraphics[width=\textwidth]{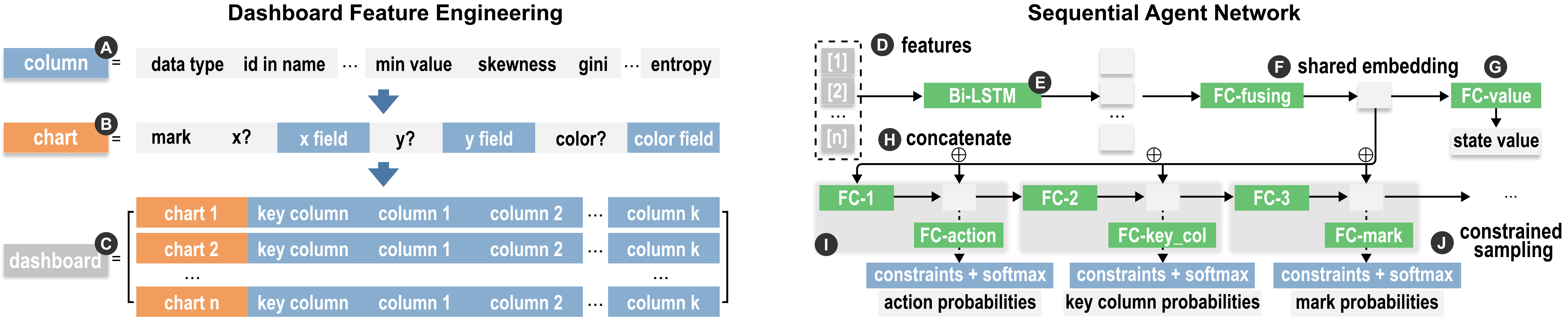}
	\vspace{\vslen}
	\caption{Process of feature construction and structure of sequence generation network. The dashboard features (C) are constructed from column features (A) and chart features (B). The dashboard features (D) are fed into the sequence generation network with a Bi-LSTM layer (E) for the extraction of shared embedding (F). The shared embedding is forwarded to a fully-connected layer for value estimation (G). The shared embedding is then sequentially fed into several classification blocks (I) and concatenated with intermediate embedding (H) for the prediction of action and parameter probabilities with constrained sampling applied (J).}
	\vspace{\vplen}
	\label{fig:network}
\end{figure*}

The asynchronous mechanism of A3C enables the training by exploration with multiple independent agents, which can accelerate the convergence of networks.
{The design of agent network comprises feature engineering of dashboards (\autoref{sec:feature}), sequential network structure for action and parameter selection (\autoref{sec:network}), and constrained sampling with visualization design rules (\autoref{sec:sampling}). The detailed structure combining these modules together is demonstrated in \autoref{fig:network}.}

\subsection{Dashboard Feature Construction}
\label{sec:feature}
The construction of a deep neural network requires numerical representations for the input data.
However, it lacks a proper feature engineering method for dashboards.
Recently, Wu et al.~\cite{wu2021multivision} proposed to model the dashboard features in an indirect way.
They firstly obtained single chart features through a learning-to-rank neural network and then packed the features together as the representation of the dashboard.
Though useful, such a method requires another network for feature engineering.
% , resulting in complex model structures.
At the same time, other studies, such as VizML~\cite{hu2019vizml} and KG4Vis~\cite{li2021kg4vis}, propose to represent data columns with hand-crafted features, which are proven to be effective through experiments.
%  and construct chart representations from the column features.
% Their experiment results demonstrate the usefulness of hand-crafted column features.
Therefore, we propose to combine the learning-based and hand-crafted features for dashboard representation. 
Specifically, we first incorporate hand-crafted column features to construct dashboard representations.
Then the representations are fed into neural networks to learn high-level embedding for action and parameter prediction.

We first construct column features from column properties and statistics metrics based on VizML~\cite{hu2019vizml} (\autoref{fig:network}A).
Column properties include data types (e.g., quantitative, nominal, or temporal),  minimum value, cardinality, etc.
Statistics metrics consist of values computed from statistics models, such as skewness and Gini impurity~\cite{wiki:gini}.

Based on the column features, we further construct chart features by representing chart attributes (\autoref{fig:network}B).
Our representation is based on Vega-Lite~\cite{arvind2017vega}, a declarative programming language popular for chart rendering.
We primarily focus on mark types of bar, line, point, and boxplot, and visual channels of x, y, and color.
The representation can be easily extended for more mark types and visual channels with a similar modeling method.
For each chart, we represent mark type and the use of visual channels with one-hot encoding.
{If a visual channel is used, we append the feature of the encoded column (denoted as field feature).
It is noted that the field feature is not the copy of the column feature, but the features of the encoded data after data transformation for rendering.
Taking \autoref{fig:case}-B1 as an example, the features of ``y'' field would be the features of ``\textit{mean US Gross grouped by Major Genre}'' instead of ``\textit{US Gross}.''}
% For example, for a histogram visualizing the distribution of a quantitative column, where x axis encodes the binning of the column and y axis encodes the count for each bin,
% the features of x field and y field are computed after the binning and counting transformation.
Hereby, we intrinsically embed the fields of ``aggregate'' for each visual channel.

The dashboard features are constructed by packing the chart features together with context information about the key column and datasets (\autoref{fig:network}C).
For each chart, we append the features of the current key column and all data columns to facilitate the column selection during exploration.
We set a maximum number of 10 for data columns.
For datasets with less than 10 columns, we pad the chart features with zeros.
Finally, we obtain the dashboard features at step $i$, $e_i\in \mathbb{R}^{n\times l}$, where $n$ is the current chart number and $l$ is the length of chart features with contextual information added.

\subsection{Agent Network Architecture}
\label{sec:network}
On the basis of dashboard features, we design an agent network for action prediction and parameter selection.
The design of the network should fulfill three requirements.
First, the network can learn mutual relationship between charts and generate a unified representation for the dashboard.
Second, the network is supposed to achieve value estimation, action prediction, and parameter selection with an integrated structure.
Besides, the network should predict parameters considering their interrelation.
For example, when adding a new chart, the agent should focus on chart configurations, but when deciding to change the key column, the agent should consider the columns to be selected.

To achieve the first requirement, we introduce a long short-term memory (LSTM) layer to model the relations between charts.
In our task, we aim to learn the chart combinations instead of sequence orders.
However, it is computationally costly to enumerate the chart combinations with all possible orders.
As a compromise, we use a bidirectional LSTM (Bi-LSTM)~\cite{huang2015bidirectional} layer that summarizes the information from both directions (\autoref{fig:network}E).
In addition, we randomly shuffle the charts during training, which generates dashboards with different chart orders.
The embedding extracted from the Bi-LSTM is then fused to be a unified shared embedding for the dashboard (\autoref{fig:network}F).

For the second requirement, we incorporate multiple classification blocks for value estimation, action prediction, and parameter prediction.
First, the value estimation is achieved with a fully connected layer taking the shared embedding as input (\autoref{fig:network}G).
Next, for action and parameter prediction, we 
% Furthermore, instead of feeding shared embedding into several independent classification heads
introduce a sequential model structure that predicts the parameters by incorporating the embedding from previous predictions of actions and parameters.
Specifically, the shared embedding is sequentially passed into several classification blocks (\autoref{fig:network}I), each block responsible for predicting an action or a parameter.
The inputs are forwarded into a fully connected layer in a classification block to obtain an intermediate embedding, which is then fused with the shared embedding (\autoref{fig:network}H) and fed into a fully-connected layer.
The fused embedding is then forwarded into the next classification block for the prediction of the following parameter.
With such a sequence model, the prediction of parameters can refer to the previous context information and thus accelerate the convergence.
In each classification block, the outputs are fed into a categorical softmax layer, and probabilities for different actions or parameters are obtained (\autoref{fig:network}J).
% The prediction results in standard classification models are obtained by selecting the one with the highest probabilities.
% However, in reinforcement learning, to make the exploration more diverse, sampling is adopted to enable the selection of actions with lower probabilities.
% Specifically, categorical sampling is performed for the actions or parameters based on the softmax probabilities.

\begin{figure*}[!htb]
	\includegraphics[width=\textwidth]{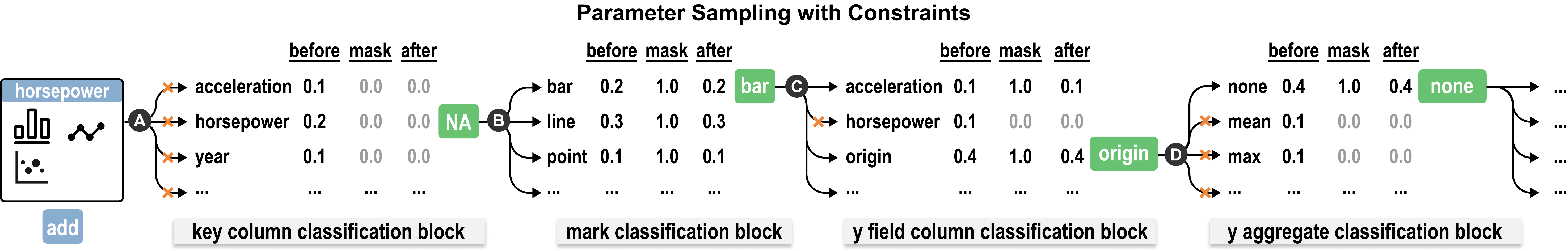}
	\vspace{\vslen}
	\caption{An example of constrained sampling for configuring a new chart. Masks are applied in the classification blocks to avoid infeasible parameter selections for key columns (A), mark types (B), y field (C), and y aggregate (D).}
	\vspace{\vplen}
	\label{fig:sampling}
\end{figure*}

\subsection{Constrained Sampling}
\label{sec:sampling}
When using an agent network for action execution with parameters, it is critical to control the activation of different network components for three reasons.
First, the column numbers of different datasets are different.
It is supposed to avoid the selection of non-existing columns.
Second, it is supposed to activate different branches of parameters based on the actions.
For example, when the agent selects to ``change'' the key column, the branches for configuring charts should be disabled.
Third, it is required to rule out improper parameter values based on the selection of previous parameters.
The parameters in chart configurations are interrelated. 
However, the agent network possibly generates incorrect chart configurations which violate the grammar of visualization rendering.
False configurations might result in failed charts without any analytical and aesthetic value.
% \dazhen{consider move to the discussion part} 
To alleviate the situation, we attempted to introduce a large penalty for invalid chart configurations.
{The agents succeeded in reducing the generation of invalid charts}, but the balancing between penalties and rewards for insights and chart designs becomes another critical problem (\autoref{sec:ablation}).
Nevertheless, even the best machine learning models might make mistakes.
Therefore, to ensure the correctness of chart generation, constraints are commonly used to regulate the model weights.
For example, to generate valid chart configurations from natural languages, NL2Vis~\cite{luo2021natural} uses constraints to mask the attention of transformer networks to avoid false inference.

Similarly, we formulate visualization knowledge as constraints to ensure sampling reasonable parameters for dashboard exploration.
Our design of constrained sampling is based on sequence modeling, in which the inference is performed sequentially. When an inference of previous action or parameter is obtained, constraints based on visualization knowledge will be applied in the next classification block to ensure the validity of the final configuration. 

We demonstrate how constrained sampling works through an example of configuring a bar chart (\autoref{fig:sampling}).
% An example showing how constrained sampling works on the generation of a bar chart configuration is demonstrated in \autoref{fig:sampling}.
In the example, the action classification block of the agent decides to ``add'' a new chart for a current dashboard with the key column of ``horsepower.''
The agent has to configure a new chart by specifying parameters.
% Given that the current dashboard has a selected topic, the ``add'' operation is not allowed to change a new topic.
Therefore, constrained sampling is applied on the key column classification block disabling all weights and generating a dummy sampling.
For mark types, all choices are available.
The agent samples the ``bar'' based on the softmax value, although the ``bar'' is not the one with the highest probability.
After selecting the mark type, the agent network predicts an explanation column.
The selection of ``horsepower'' is disabled because ``horsepower'' is the key column.
The agent selects ``origin,'' which is a nominal column, so the ``aggregate'' of mean and max are disabled.
Next, the agent will generate an ``aggregate'' of the key column and color encodings with similar behaviors.
With such an activation mechanism, we ensure the validity of the sampled configurations.
It is noted that the constraints are applied to the predictions before softmax, which ensures the validity of entropy computation for back-propagation.

\subsection{\system{} Interface \& Interactions}
To facilitate interactive creation, we design an interface consisting of a table view (\autoref{fig:teaser}A), a topic view (\autoref{fig:teaser}B), a chart editor (\autoref{fig:teaser}C), a canvas view (\autoref{fig:teaser}D), and a recommendation view (\autoref{fig:teaser}E).

The table view supports uploading a CSV data file for dashboard generation.
After uploading, the columns and their types are shown in a list.
% Users can modify column type if the results of automatic parsing are incorrect.
Users can view the raw data in a table visualization by clicking the view button.
{Meanwhile, the agent network automatically explores and generates dashboards directly.
Specifically, an agent is allowed to explore at most 50 steps for a dashboard in case of no early termination (e.g., the number of charts reaching the maximum limit).
We assign the agents a quota of $n$ steps ($n=1000$), and the agents can generate a series of dashboards with different topics (i.e., key columns).}
The dashboards with different topics will be displayed in the topic view sorted by returns.
Users can drill down to investigate each dashboard.
{When hovering on another result, a tooltip will pop up demonstrating the changes on charts and insights.}

The canvas view displays the charts generated by the deep reinforcement model.
Displaying multiple views of charts is a critical challenge for visual analysis and has been studied for a long~\cite{chen2020composition}.
% Different from data stories in which there is a narrative order between charts, all charts in a dashboard have interrelated connections.
Given that chart layout is not the research focus of this work, we use a rule-based method to provide an efficient solution.
Specifically, we first aggregate the charts by their mark type. For the charts with the same mark types, we aggregate the ones with the same insight types.
From the preliminary study, we understand that text visualizations are commonly used to show an overview of the data.
Therefore, we summarize the statistics of key columns and other columns that are possibly interesting and visualize the statistics by a default setting of text visualizations.
The text visualizations will be positioned in the top row of the dashboard.
The chart sizes and positions are also changeable.

{With chart editor, users can edit the dashboards with online recommendations.
Users are allowed to edit chart configurations by editing the parameter values, adding a new chart, or deleting charts.
Once edited, the agents will start exploring from the current dashboard state for k steps (k=200).
The systems would identify the best charts that could be added to the dashboard and maximize rewards.}
The recommended charts will be shown in the recommendation view.
Users can explore and add the ones of interest to the dashboard.

\begin{figure}[tb]
	\includegraphics[width=\linewidth]{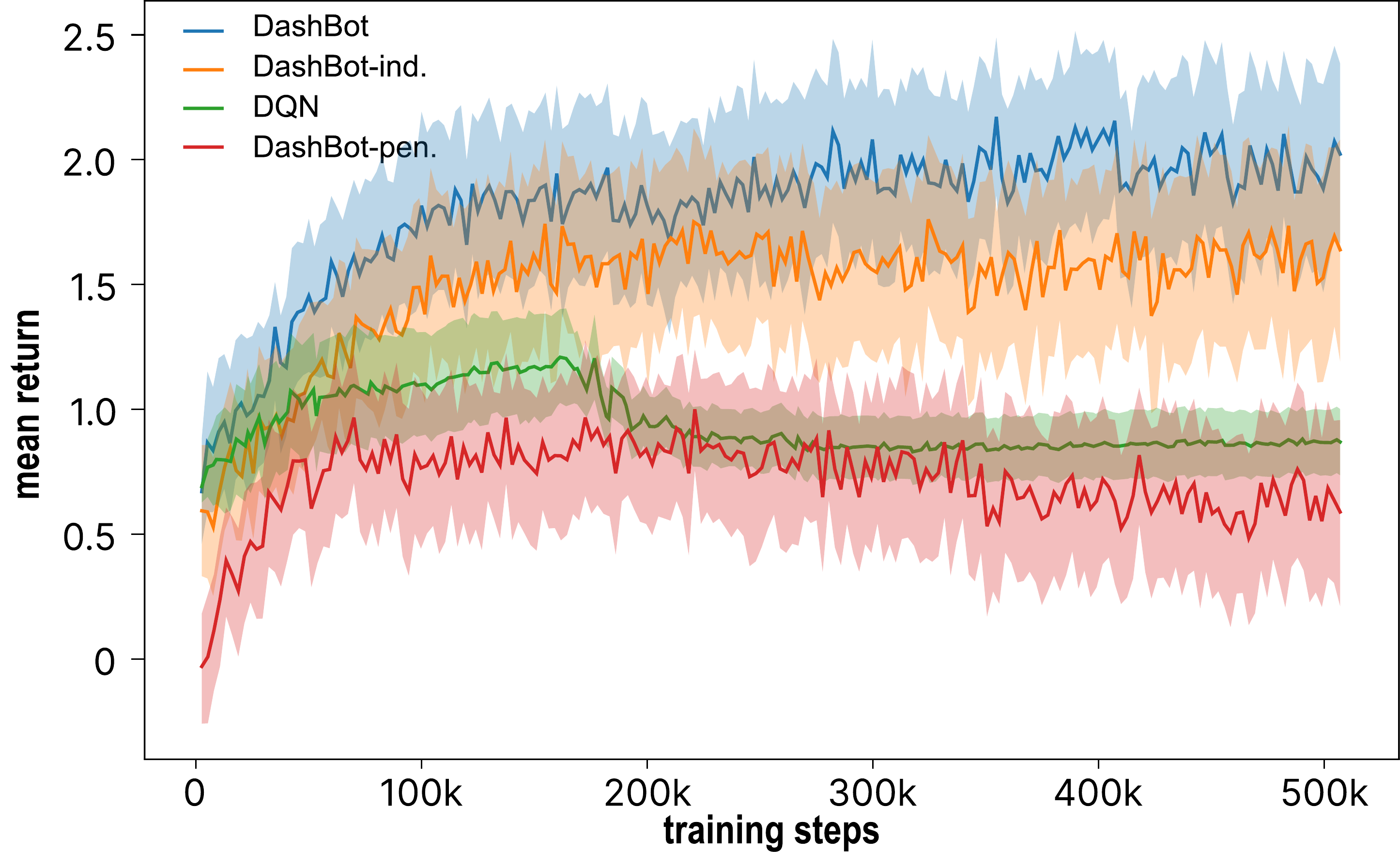}
	\vspace{\vslen}
	\caption{Ablation study showing the mean returns of \system{}, \system{}-ind., \system{}-pen., and DQN across training steps.}
	\vspace{\vplen}
	\label{fig:ablation}
\end{figure}
\begin{figure*}[!htb]
	\includegraphics[width=\linewidth]{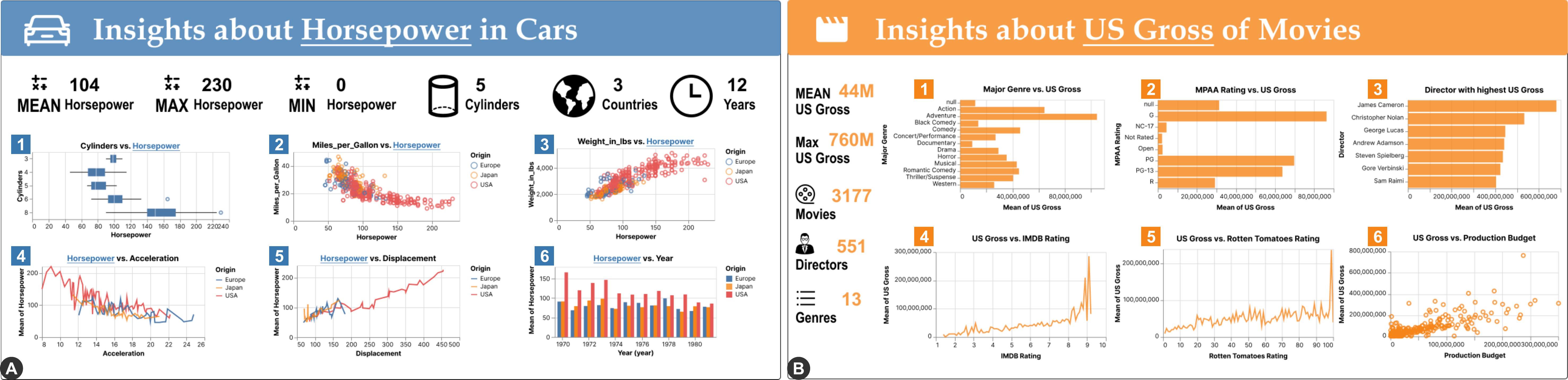}
	\vspace{\vslen}
	\caption{Two example Dashboards created with \system{}: (A) a dashboard showing the properties of cars, such as cylinders, miles per gallon, and accelerations, about horsepower and (B) a dashboard about US gross of the movies with regard to IMDB ratings, Rotten Tomato ratings, genres, etc.}
	\vspace{\vplen}
	\label{fig:case}
\end{figure*}
\section{Evaluation}
To demonstrate the usefulness of \system{}, we first show three example dashboards created with \system{} (\autoref{fig:teaser}, \autoref{fig:case}).
Second, we focus on evaluating the core component of \system{}, i.e., the deep reinforcement learning model that defines the dashboard design space, enumerates that space, and ranks the candidate designs~\cite{zeng2021evaluation}, through an ablation study and a user study.
% Furthermore, to demonstrate the usefulness of the overall interface, we conduct an expert interview with senior visual analytics experts.

\subsection{Example Dashboards}
We first show how to exploratory generate a dashboard through the first example (\autoref{fig:teaser}).
{Then we show two automatically generated examples (\autoref{fig:case}) with style redesigned on the exported SVG files using Figma.}

In the first example (\autoref{fig:teaser}), assume that Mary wants to investigate the insights into Seattle weather.
After uploading the data file, the data columns and types are shown (\autoref{fig:teaser}A).
Then a series of dashboards are generated by \system{} and listed in the topic list (\autoref{fig:teaser}B).
Mary is more interested in the topic of ``wind'' in Seattle.
Therefore, she selects the dashboard of ``wind'' with the highest return value (\autoref{fig:teaser}D).
From Chart 1, she discovers that the minimum temperature seems to have no obvious correlation with the wind, but there is more red points on the right side.
She infers that the days with the highest winds are mostly rainy.
Chart 2 shows wind distribution by different weathers.
The proportion of rainy days gets larger as the wind gets larger.
This pattern partially conforms with the inference in Chart 1.
Chart 3 provides more evidence with a boxplot visualization.
The median wind on rainy days is higher than on drizzling, foggy and sunny days, but lower than on snowy days.
Furthermore, the points on the rightmost indicate that days with the largest wind are all rainy.
Chart 4 presents the wind change across dates.
Mary further feels interested in the dates with the highest winds and thus configures a new chart, Chart 5, with chart editor (\autoref{fig:teaser}C).
After the edition, \system{} automatically recommends Chart 6 about the days of lowest wind for comparison.

The second example (\autoref{fig:case}A) is about the horsepower in the ``cars'' dataset.
Chart 1 is a boxplot showing horsepower distribution by cylinders.
Charts 2-5 indicate that miles per gallon and accelerations are possibly negatively correlated with horsepower while weights and displacements tend to be positively correlated with horsepower.
Chart 6 further gives an overview of horsepower with regard to years.

The third (\autoref{fig:case}B) is about US gross in the ``movies'' dataset.
Chart 1 \& 2 demonstrate the mean US gross by major genres and MPAA ratings.
Chart 3 lists the directors with the highest US gross.
Chart 4 \& 5 show the US gross by IMDB ratings and Rotten Tomato ratings, which facilitate the comparison of different rating systems.
Chart 6 presents the correlation between US gross and production budget.

\subsection{Ablation Study}
\label{sec:ablation}
We opt to demonstrate the effectiveness of our model design, including the A3C framework, sequence modeling, and constrained sampling.
We design experiments with three baseline models for the ablation study.
The experiments were carried out on a workstation with 6 core Intel i7-8700K CPU, a GeForce RTX 1080Ti GPU, and 32GB memory.
All models were trained for 500,000 steps with 27 datasets from Vega Datasets\footnote{https://github.com/vega/vega-datasets}.
As suggested by Minh et al.~\cite{mnih2016asynchronous}, we train all models for 10 runs and average the return values. The results are shown in \autoref{fig:ablation}, where transparent bands represent standard variances of the return.

\textbf{A3C versus Deep Q-Network.} To demonstrate the effectiveness of the A3C training framework for the task of dashboard generation, we design a multi-step deep Q-Network (i.e., DQN) as a baseline.
The DQN uses a similar agent network as \system{}.
The DQN is trained with a temporal difference and replay memory~\cite{mnih2013playing}.
From \autoref{fig:ablation}, we discover that DQN fails to achieve a stable increase in the return.

\textbf{Sequence Model versus Independent Multi-Classification.} We design a baseline model (i.e., \system{}-ind.) by removing the embedding concatenation in classification blocks.
All classification blocks independently predict parameter values with the shared embedding. 
To ensure a fair comparison, constrained sampling is preserved to avoid invalid configurations.
From the results (\autoref{fig:ablation}), we can see the return of \system{}-ind. constantly increases but finally converges at a mean return value lower than \system{}.
The \system{} also converges faster with smaller training steps. The reason might be that the sequence model is able to learn mutual relations between consecutive tokens.

\textbf{Constrained Sampling versus Invalidity Penalty.} We design a baseline model without constrained sampling (i.e., \system{}-pen.). Alternatively, we add penalty rewards. Specifically, if an agent configures an invalid chart, it will receive a negative reward.
The penalty rewards result in a relatively low return at the beginning because of the random sampling of invalid configurations.
As the training proceeds, the agents obtain increasing returns, but finally, the return gets decreased.
The penalty mechanism seemingly works, but the balancing between penalties and other rewards might be a critical problem.

In addition, we run the \system{} for 1,000 steps (about 20 episodes) on each dataset, which takes 15.4 (SD=1.52) seconds on average, and compute the statistics of generated dashboards.
On average, the dashboards contain 5.42 (SD = 0.83) charts with 2.81 (SD = 0.74) chart types, which are consistent with the statistics in the preliminary study.

\subsection{Comparative Experiment}
We further conduct a user study to evaluate the usefulness of our deep reinforcement learning-based method in comparison with MultiVision~\cite{wu2021multivision}, another deep learning-based method.
{Two systems have different interface designs and implementation details, but the machine learning models have the same inputs and outputs.
The study focuses on evaluating the core machine learning components of the systems.}

\textbf{Participants.} We conducted the study with 10 data workers (P1-P10, 4 females and 6 males) with diverse backgrounds, including statistics, chemistry, biology, environment engineering, transportation management, computer science, finance, and economics.
All participants reported having more than two years of using programming languages (i.e., Python, R, MatLab, and Javascript) or interactive tools (i.e., Stata, SPSS, Origin, and Excel) to process and visualize data for analysis.
All participants mentioned that they only had a basic knowledge of creating charts for analysis or presentation.

\textbf{Data.} To ensure fairness, we use the same datasets in MultiVision from Vega Datasets, which include data about cars, jobs, penguins, Seattle weather, and movies.
% , which are popular in the fields of data mining and visual analytics. 
For each dataset, two systems generate one dashboard. Given that \system{} can generate multiple dashboards with different topics, we select the one with the highest return values.

\textbf{Experiment Setup.} We design a comparative experiment to evaluate the initial results generated by two systems because the two systems have provided different interaction designs for the dashboard edition and human inputs are hard to evaluate.
To ensure the fairness of comparison, the results generated by the two methods are all rendered with \system{} interface, all facilitated with text visualizations for dataset overview.
This makes sure the same styles of charts and interfaces.
{Please note that the dashboards generated with the original MultiVision interface are facilitated with interactivity (i.e., cross-filtering), but the interactivity is not considered in this study because we focus on the model comparison.}
In the study, users could explore two dashboards freely without knowing how the dashboards are generated.
We counterbalanced the presentation order in order to alleviate the carryover effect.
We followed the think-aloud protocol to collect feedback about the generated dashboards.
Specifically, during exploration, participants were requested to report insights she/he gains from the dashboards.
After reading the dashboards of each dataset, the participants were asked to select the better one with regard to overall quality, insightfulness, understandability, and aesthetics.
The experiment was followed with a post-study interview to understand participants' routine analysis workflow and suggestions about the dashboards.

\textbf{Procedure.} A study lasted for about 50 minutes with a 10-minute training session, a 30-minute experiment, and a 10-minute post-study interview. Before the study, participants were asked to sign a consent form agreeing to record their feedback and comments for analysis.
All participants were paid with \$20 after the study.

\begin{figure}[tb]
	\includegraphics[width=\linewidth]{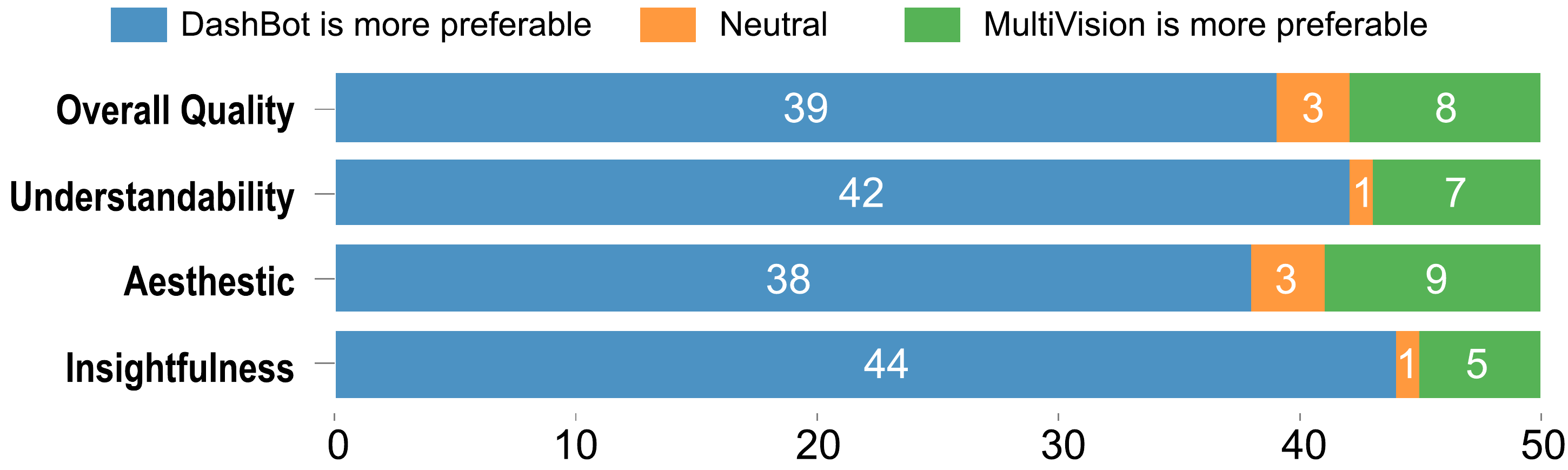}
	\vspace{\vslen}
	\caption{Ratings of \system{} compared with MultiVision from the perspectives of overall quality, understandability, aesthetic, and insightfulness.}
	\vspace{\vplen}
	\label{fig:user}
\end{figure}

\textbf{Results.} In total, we collected 200 ratings from 10 participants.
As shown in \autoref{fig:user}, 78\% of ratings agree that dashboards generated by \system{} have higher overall quality when compared with the baseline method. 
We collected a large number of positive comments about our method and summarized the comments below.

Among the responses, 84\% told that \system{} is more understandable.
Participants liked our idea of key columns.
P8 quickly identified that the charts in our generated dashboard share a common column, saying that ``\textit{the design accelerates the understanding of the dashboard because I only need to identify the columns combined with that (key) column}'' (C1).
Due to the constrained sampling to avoid ineffective encodings, the dashboards generated by \system{} might be easier to understand.
P6 appreciated the use of color encodings on the nominal columns with an adequate cardinality and commented that ``\textit{the color is efficient for identifying different penguin species}'' (C2).

Moreover, 76\% of voters agreed that \system{} is more aesthetically pleasing.
Participants liked the diversity of chart types in the dashboards generated by \system{}.
P7, who had a background in digital media design, commented that ``\textit{the first one (our method) looks good because of diverse chart types, while the second one contains too many scatterplots}'' (C3).
P6 also held a similar opinion.
Reasonable visual encodings also improve the overall appearance.
P4, who reported to have little experience in design, raised a comment that ``\textit{this dashboard looks more comfortable at my first glance because it does not use size to represent the column of Miles\_per\_Gallon (in the scatterplot)}'' (C4).
P1 liked the aggregation of columns, which benefits from the design of an agent network that covers different chart parameters, saying that ``\textit{averaging the values of this column makes the chart clear}'' (C5).

More participants (88\%) thought that \system{} can provide more insights compared with the baseline.
P6 commented that ``\textit{the second one (our method) provides more column combinations other than the other one, which is more informative}'' (C6).
P5 liked the patterns exhibited in the dashboards we generated, saying that ``\textit{I can easily identify the correlations from the shape of scatterplots}'' (C7).
P2 inferred additional column relationships from the charts, commenting that ``\textit{these two scatterplots (with high correlation values) potentially reveal the relations between another two columns, although they are not visualized in the same charts}'' (C8).
P4 showed her preference for the comparative analysis on the top and bottom entries, with a comment that ``\textit{if I were a climate analyst, showing me the dates with highest and lowest temperatures will be meaningful}'' (C9).

{\textbf{Analysis.} We summarized participants' feedback and analyzed the technical differences that theoretically make \system{} outperforms MultiVision. 
First, numerically modeling insights helps to reveal important information about the data.
A major difference between \system{} and MultiVision is that we evaluate the insightfulness of the charts with computational metrics. Thus, the recommended charts have data patterns that are more salient and easier to comprehend (C7, C8).
Second, incorporating dashboard design patterns into the generation is another advantage of \system{}. For example, we require the charts in a dashboard to have a shared column as the topic of the dashboard, which makes the results more coherent and interconnected (C1).
MultiVision incorporates criteria such as diversity and simplicity into the dashboard design.
\system{} follows a similar approach, but further considers statistics obtained from empirical surveys to follow ``common practices'' (C3, C6).
Third, constrained sampling help exclude charts with ineffective visual encodings.
MultiVision recommends chart encodings by learning from the Excel datasets, which might result in some unreasonable recommendations. For example, MultiVision tends to visualize three data columns in a chart and encodes numerical data with point size, which is complained about by the participant (C4). Instead, \system{} takes visual effectiveness into the design of constrained sampling for better visual encodings and data transformation (C2, C5).}

\section{Discussion and Limitations}
In this section, we discuss our limitations and potential improvements.

\subsection{Extending Reward Function Sets}
{We design reward functions based on a preliminary study of dashboard designs, but there are additional considerations to extend the reward function set.
For presentation, we only consider the diversity and parsimony between charts. Chart effectiveness is also an important aspect of chart comprehension~\cite{lo2022misinformed,chen2021vizlinter}.
A potential solution is introducing Draco knowledge base~\cite{moritz2018formalizing}, a comprehensive rule set of visualization design knowledge, to compute the effectiveness scores of charts and integrate them into reward functions. Moreover, the effectiveness of chart composition~\cite{javed2012composite, deng2022revisiting} should be considered. For example, even with a large number of subplots, scatterplot matrices are still easy to understand. Thus, continued research on quantifying the quality of chart composition is needed.}
%  Effective chart compositions may receive a large reward.

{For insights, we use several metrics (e.g. correlation) to measure the insightfulness of a chart based on existing studies~\cite{tang2017extracting, shi2020calliope}. However, our current reward functions do not consider semantic information in the columns, which might lead to results of commonsense~\cite{karer2021insight}.
For example, the total sales of a merchant in a year are certainly correlated to the average sales.
A potential fix for this problem is to process the header names for their semantic structure~\cite{du2021tabularnet} for reward function design (e.g., normalizing the correlation with semantic distance).
The introduction of semantic information is potentially helpful for addressing domain problems.
For example, when exploring the dashboards of the penguin dataset, P10, who has a background in biology, commented that comparing statistics between different species would be extremely important, and using y positions might be more intuitive than colors.
The environment should reward the agent if a chart visualizes a topic-related column with an effective visual channel.
% In addition, some metrics we use might be not insightful enough, such as ``distribution,'' and more in-depth measurements for the distributions are required.

Additional metrics could be incorporated into reward functions for better insight discovery.
For example, as suggested by participants with a statistics background (P3 and P9), Shapiro-Wilk static, skewness, and kurtosis can be used to evaluate the normality of the data distribution.
Visual patterns are another critical consideration for insight discovery by human experts.
% For example, P8 immediately identified that the percentage of men and women workers becomes more and more balanced across the years, which conforms to her knowledge.
Computer vision~\cite{ma2018scatternet} and other sophisticated algorithms~\cite{behrisch2018pattern,li2022structure} could be used to measure the visual salience of the patterns, such as outliers and visual grouping~\cite{giovannangeli2022color, wang2022makes}.
}

\subsection{Evaluating \system{} in Real-World Scenarios}
{It lacks comparisons between \system{} and what users might do with real-world tasks.
User performance in different recommendation systems is a critical aspect of evaluation~\cite{van2013evaluation, zeng2021evaluation}.
However, directly comparing two complex systems is difficult in controlling dependent variables (e.g., workflows, interface designs, and artifacts)~\cite{lazar2017research}.
In our case, \system{} has a different workflow and interface design.
% In our case, \system{} has a different workflow and interface designs, which might introduce biases.
% First, different workflows make it challenging to measure the time efficiency. 
\system{} recommends a default setting and allows users to adjust the dashboards to their satisfaction, while real-world tools (e.g., Tableau and Power BI) provide efficient interactions to support generating dashboards from scratch.
Moreover, different dashboard styles might introduce biases in interpreting the insightfulness and understandability of the results.
\system{} is a proof-of-concept system that only covers the generation of dashboards with limited chart types and layouts and lacks support in style customization.
In contrast, real-world authoring tools support the generation of sophisticated dashboards even with advanced pictograms.
Controlled studies of multiple recommendation systems could be conducted with a consistent interface to evaluate user performance~\cite{zeng2021evaluation}.
Future evaluation could consider integrating the \system{} model into existing tools and compare how the resulting dashboards are different under different workflows.

Real-world tasks also differ from lab studies in terms of datasets.
We use Vega datasets throughout the paper, given that the datasets are commonly used in visualization generation~\cite{wongsuphasawat2017voyager, narechania2020nl4dv} and the area of data analysis, which is understandable for readers and study participants.
It is also convenient for model comparison because the baseline model was trained on Vega datasets.
However, real-world datasets could contain hundreds or thousands of data columns, which challenges the scalability of models.
Moreover, additional table operations (e.g., reshape and rearrange) should be considered for data cleansing and insight discovery~\cite{kasica2021scraps}.
Therefore, it would be a total different problem for feature engineering, model design, and the overall framework design (e.g., action space) when dealing with real-world datasets.
Future study could focus on developing efficient agents to explore and identify insights from large-scale datasets.
}

\subsection{Considering Additional Data Transformation}
\system{} currently considers basic data aggregation calculators (e.g., mean and sum) and sorting for identifying the top and bottom values.
% Echoed by P1, additional operations such as filtering can be considered.
P9 suggested encoding multiple data series in the same chart by introducing visualization compositions such as layering and faceting.
Given that the visualization rendering of \system{} is achieved using Vega-Lite, these potential operations can be extended with additional modeling of Vega-Lite parameter space.
Given that we primarily focus on specifying visual encodings, some chart appearances are not aesthetically pleasing. For example, the line chart of \autoref{fig:teaser}-D4 is too dense for investigation, requiring binning or sampling. 
% P2 noted that scaling operations should be considered to improve the visual appearance.
% Providing an interface to edit Vega-Lite configurations could be a straightforward solution.
Towards the generation of delicate charts, it would be a promising research direction to encode the knowledge of visualization debugging~\cite{chen2021vizlinter} and layout optimization~\cite{wu2020mobilevisfixer} into the deep reinforcement learning framework.

\subsection{Enhancing Dashboards with Interactivity}
{\system{} focuses on modeling data insights and chart configurations in dashboards but does not support interactivity, which limits the usability of the generated dashboards~\cite{sarikaya2018we}.
Bridging the charts with interactivity requires the modeling of data flow between charts.
MultiVision~\cite{wu2021multivision} supports cross-chart filtering by detecting the data records related to user's filtering operations on a single chart.
This solution is efficient when all charts share the same datasheet.
However, there might be more complex situations.
For example, the data of a chart might be transformed from the data of another chart.
To handle this situation, the data flow model should be constructed when the exploration and data transformation are executed, as demonstrated by VisFlow~\cite{yu2016visflow}.
Such kind of progressive exploration could be integrated into the framework of \system{}, where agents could choose to explore from the data of the previous chart or use the original table when generating a new chart.
Reward functions that model the relationship between the consecutive charts should be considered~\cite{kim2017graphscape}.}

% \textbf{Limited Support for Dashboard Styling.}
% Currently \system{} supports modifying chart sizes and positions, but does not support the customization of overall styles of the dashboards.
% To further enhance the visual styles, users have to download the SVG files generated by \system{} and edit the styles with graphics design tools.
% As suggested by P2, we can consider the topic of datasets and recommend relative color palettes for the color encoding and dashboard styles.

\section{conclusion and future work}
We present \system{}, a deep reinforcement learning model, to generate analytical dashboards for insight discovery.
To design an effective model, we conduct a preliminary study to understand the design practices in existing dashboard galleries.
Then we formulate the problem of exploratory dashboard generation as a Markov decision process with appropriate action spaces and reward functions.
Furthermore, we design a sequence generation network for the action selection and parameter configuration.
Finally, we demonstrate the effectiveness of our model design through an ablation study and a user study.
Our study opens up a research direction that facilitates visualization generation without the need for large-scale human-labeling training data, which is a pain point for learning-based visualization generation methods.

Further research can be conducted in two directions. First, we can extend the framework for the generation of more visualization genres, such as glyph~\cite{ying2021glyphcreator,ying2022metaglyph} and data stories~\cite{shu2020makes}, considering additional visualization types and data transformation operations.
Moreover, the data flow could be considered to improve the interactivity.
Second, we can incorporate the visual patterns into the modeling of insight discovery~\cite{ma2018scatternet}, given that visual patterns are unique values that visual analytics can provide for expert users.
Third, inspired by the idea of inverse reinforcement learning~\cite{ng2000algorithms}, we can further incorporate user data to steer the generation of dashboards on the basis of visualization rules. User inputs can guide the agents to adapt to new analysis scenarios.
%% if specified like this the section will be committed in review mode
\acknowledgments{
The work was supported by NSFC (62072400) and the Collaborative Innovation Center of Artificial Intelligence by MOE and Zhejiang Provincial Government (ZJU).}

\end{spacing}

\bibliographystyle{abbrv-doi}

\bibliography{main}
\end{document}